\documentclass[aps,prl,twocolumn,groupedaddress,amsmath,amssymb,nofootinbib,showpacs,showkeys]{revtex4-1}
\usepackage{graphicx}  
\usepackage{dcolumn}   
\usepackage{color}
\usepackage{psfrag}
\usepackage{bbm,bm}
\usepackage{epsfig}
\usepackage{hyperref}
\allowdisplaybreaks
\begin{document}
\def\be{\begin{equation}}
\def\ee{\end{equation}}
\def\bea{\begin{eqnarray}}
\def\eea{\end{eqnarray}}
\def\ba{\begin{array}}
\def\ea{\end{array}}
\def\ben{\begin{enumerate}}
\def\een{\end{enumerate}}
\def\nab{\bigtriangledown}
\def\tpi{\tilde\Phi}
\def\nnu{\nonumber}
\newcommand{\eqn}[1]{(\ref{#1})}
\def\bw{\begin{widetext}}
\def\ew{\end{widetext}}
\newcommand{\half}{{\frac{1}{2}}}
\newcommand{\vs}[1]{\vspace{#1 mm}}
\newcommand{\dsl}{\pa \kern-0.5em /} 
\def\a{\alpha}
\def\b{\beta}
\def\g{\gamma}\def\G{\Gamma}
\def\d{\delta}\def\D{\Delta}
\def\ep{\epsilon}
\def\et{\eta}
\def\z{\zeta}
\def\t{\theta}\def\T{\Theta}
\def\l{\lambda}\def\L{\Lambda}
\def\m{\mu}
\def\f{\phi}\def\F{\Phi}
\def\n{\nu}
\def\p{\psi}\def\P{\Psi}
\def\r{\rho}
\def\s{\sigma}\def\S{\Sigma}
\def\ta{\tau}
\def\x{\chi}
\def\o{\omega}\def\O{\Omega}
\def\k{u}
\def\pa {\partial}
\def\ov{\over}
\def\nn{\nonumber\\}
\def\ud{\underline}
\def\ct{\textcolor{red}{\it cite }}



\title{\large{\bf Finite temperature $0^{-+}$ glueball spectrum from non-susy D3 brane \\ of Type IIB string theory}}
\author{Kuntal Nayek}
\email{kuntal.nayek@iitkgp.ac.in}
\affiliation{
   Indian Institute of Technology,\nn
   Kharagpur, West Bengal 721302, India\nn
   }
\author{Shibaji Roy}
\email{shibaji.roy@saha.ac.in}
\affiliation{
   Saha Institute of Nuclear Physics,\nn
   1/AF Bidhannagar, Kolkata 700064, India\nn
   and\nn
   Homi Bhabha National Institute,\nn
   Training School Complex, Anushakti Nagar, Mumbai 400085, India
   }

\begin{abstract}
Here, we calculate the pseudo-scalar glueball mass at finite temperature from the holographic 
QCD in $3+1$ dimensions. The decoupled geometry of the non-supersymmetric (non-susy) D$3$
brane at finite temperature, which is the solution of type IIB 
supergravity, is considered as the dual theory of QCD. We calculate the mass 
spectrum from the axion fluctuation in this gravity background using WKB 
approximations. Approximating the WKB equation for various orders of mass, we 
derive the analytical expressions of the mass spectrum of $0^{-+}$ at 
finite temperature. Finally, we evaluate the masses of the ground state
and the first excited state of the pseudo-scalar glueball numerically, as, $m_{-+}=2.5$GeV and 
$m_{-+}^*=3.8$GeV respectively, from the complete WKB equation. The mass of a given state 
is found to decrease with increasing temperature and becomes zero at confinement-deconfinement
transition temperature which is consistent with the idea of 
confinement and also matches with some recent lattice results. From this 
temperature variation, the QCD transition point is found to be about 
$186$MeV.
\end{abstract}


\maketitle

The Relativistic Heavy Ion Collider (RHIC) has already 
confirmed the plasma state of nucleons. At extremely high energy, the nucleons 
completely dissolve and produce the plasma state of free quarks and 
gluons without interactions. As energy decreases we move 
from the deconfined (plasma) state to confined (bound state of quarks and 
gluons) state and at sufficiently low energy, we get back the nuclear 
matters or the hadronic states without any free quarks and gluons. Now when the energy 
decreases from the plasma state the interactions among the constituents dominates and 
the interacting gluons form glueballs. Thus we get a mixed phase of 
free gluons and glueballs. As the energy decreases further the glueball 
mass increases gradually as a result of increasing coupling. In this 
transition process, we have a particular range of temperature ~$175\pm 15$MeV, 
below which we have the glueball dominated phase or the confined phase and 
above which we have the free gluon dominated phase or the deconfined phase. 
So, the glueball spectrum is essential to study various important properties 
of the confined QCD. Till date, no high energy experiment has confirmed the 
existence of glueball, so there is a lack of phenomenological evidence. 
However, the MIT bag model, the Skyrme model, chiral perturbation theory, 
heavy baryon perturbation theory have been modeled to describe the confined 
phase of QCD; the glueball is not fully explored in that respect. 
The MIT bag model \cite{Jaffe:1975fd} is a 
theoretical model of the glueball for various quantum numbers (spin, 
parity etc). According to this model, depending on the number and the properties of the 
constituent single-gluon states, we can have glueballs of different spin and 
parity $J^{PC}$, where $J$ denotes spin of the glueball and $P$ and $C$ denote 
the spatial parity and the charge conjugation parity respectively. For example, $0^{++},\,2^{++},\,
0^{-+},\,2^{-+}$ are made of two gluon modes, whereas, $0^{++},\,1^{+-},\,3^{+-}$ consist
of three gluon modes. The pseudo-scalar glueball has 
spin $J=0$, parity $P=-1$ and $C=+1$ since it is chargeless. $0^{-+}$ 
consists of a transverse electric and a transverse magnetic gluon modes \cite{Mathieu:2008me}. 

In recent times, the lattice approach \cite{Lepage:1992xa,Bernard:2001av} 
is very popular to deal with the confined regime of QCD. The lattice calculations 
have found the glueball spectrum in various dimensions for both the 
quenched and the unquenched QCD \cite{Rajagopal:1992qz,Mathieu:2008me,Gregory:2012hu,
Chen:2005mg,Morningstar:1999rf}. They have also calculated various meson spectrum, 
scattering amplitude \cite{Dudek:2012xn} etc., although the lattice approach has 
some technical limitations like the finiteness and discreteness of 
the lattice and is not suitable to study the dynamics. In the late $90$s, 
after the formulation of the AdS/CFT correspondence \cite{Maldacena:1997re, 
Aharony:1999ti}, we have seen various strongly coupled gauge theories can be 
studied from holographic gravity background. In string theory, the AdS/CFT 
correspondence is a strong/weak duality with respect to the coupling constant of 
the corresponding theories. The strongly coupled gauge theory corresponds to weakly 
coupled supergravity. So when we study holographic QCD in gauge/string duality, 
we always deal with strongly coupled gauge theory. Thus the asymptotic freedom
is not accessible \cite{Polchinski:2001tt} from 
the holographic method. But other properties of QCD like confinement, RG flow, 
gluon condensate \cite{Constable:1999ch,Babington:2003vm,Csaki:2006ji} and various 
properties of QGP \cite{Liu:2006he,CasalderreySolana:2011us,Chakraborty:2017wdh} 
can be studied from dual gravity backgrounds. So a proper holographic approach for low 
energy QCD is quite significant to study the high energy particle physics and also is a
non-trivial application of string theory.  

As the effective gauge coupling in gauge/gravity duality is $\lambda=
4\pi Ne^{\Phi}$, where $\Phi$ is the dilaton field in gravity theory, therefore, to 
get the running coupling in gauge theory, the dual gravity background must have 
a non-constant dilaton field. Also the $d+1$ dimensional gravity background should not have the 
$SO(d,2)$ symmetry, so that the corresponding $(d-1)+1$ dimensional gauge theory is non-conformal 
which ensures the existence of a $\Lambda_\text{QCD}$-like fixed point. 
Considering these two properties of QCD, there are a few phenomenological models to study 
the glueballs in the holographic QCD \cite{Brodsky:2014yha,Braga:2017apr,Karch:2006pv,
Csaki:2006ji}. But those models are not directly connected to the ten dimensional 
supergravity or superstring theory. However there are some holographic QCD model 
which have been derived from the higher dimensional brane by dimensional 
compactification\cite{Witten:1998zw,Sakai:2004cn}. In this article, we consider 
the finite temperature, non-supersymmetric D$3$ brane solution of type IIB 
superstring theory. In the decoupling limit, i.e., in the low energy limit, it gives a
non-AdS geometry (however, this geometry is still asymptotically AdS) at finite 
temperature and contains a non-constant dilaton field. The dilaton field also depends 
on temperature parameter which justifies the temperature dependence of $\lambda$. On 
the gauge theory side, the pseudo-scalar glueball mass is driven by the field operator 
$\text{Tr}(F \tilde F)$ which is called the pseudo-scalar glueball field operator in 
the gauge theory. Thus the associated mass can be found from $\langle F^{\mu\nu}_a\tilde F^a_{\mu\nu}\rangle$. 
On the other hand, the pseudo scalar glueball operator couples 
to the axion field in the bulk theory. So the fluctuation of the bulk axion field gives 
the mass spectrum of $0^{-+}$. In this article we calculate the spectrum at the finite 
temperature. Here we take the perturbation of the axion field at finite temperature 
non-susy background which gives a Schr\"odinger-like wave equation. From that equation 
we evaluate the mass spectrum using semi-classical WKB approximation. The complete 
analytic solution of the spectrum is not possible due to the complicated form 
of the WKB equation. We approximate the equations for some particular 
order of mass and derive the analytic form of the temperature dependence of the 
spectra. Finally, we numerically evaluate the temperature dependent spectra for 
the full WKB equation. Here we find the glueball masses in units of 
$u_3/L^2$ where $u_3$ is the fixed mass scale of the theory and $L^4=g_\text{YM}^2N$ is 
the gauge coupling. But if we take the ratio of masses of two levels, it becomes a 
pure number. So it will be more convenient to compare the mass ratios with various 
lattice calculations rather than comparing the dimensionful mass values. Here we will 
compare the ratio of the masses of the first excited state to the ground state. On the 
other hand, by matching the dimensionful mass of a particular level with the lattice 
result at zero temperature, $u_3/L^2$ will be determined and using that value we will evaluate the masses 
of various levels in units of eV at finite temperature. Thus the full spectrum will
be drawn at finite temperature.

We have seen that the gauge/gravity correspondence is applicable for non-susy D3 brane 
solutions of type IIB superstring theory, which can be termed as 
`non-supersymmetric AdS/CFT duality' \cite{Nayek:2015tta}. The decoupled form of the finite 
temperature non-susy D$3$ brane has been discussed in \cite{Nayek:2016hsi}. The decoupled 
geometry in string frame is given in eqn.(14) of ref.\cite{Chakraborty:2017wdh} which has three 
independent parameters. Now for the following study, we take that decoupled 
background in the Einstein's frame with $\alpha+\beta=2$ which can be written 
as,
\bea
ds_E^2 & = & \frac{u^2}{L^2}G(u)^{\frac{1}{4}-\frac{\d}{8}}\left(-G(u)^\frac{\d}{2}
dt^2+d{\vec x}^2\right)\nn
 &&\quad +\frac{L^2}{u^2G(u)}du^2+L^2d\Omega_5^2\nn
e^{2\Phi} &=& e^{2\Phi_0}G(u)^{\half\sqrt{6-\frac{3}{2}\d^2}}\nn
G(u) &=& 1+\frac{u_3^4}{u^4}\label{bground}
\eea
where $-2\le\d\le0$ and $u_3$ is a fixed energy scale. Here the radius of the 
transverse sphere $S^5$ is constant. So the dimensional reduction and 
calculations become simpler. $\Phi$ is the dilaton field with the vacuum expectation 
value $\Phi_0=\ln g_s$ where the string coupling $g_s$ is related to Yang-Mills coupling
by $g_\text{YM}^2=4\pi g_s$. The gauge theory
temperature $T$ is related to the gravity theory as \cite{Kim:2007qk}
\be\label{temp}
T=\left(-\frac{\d}{2}\right)^{1/4}\frac{u_3}{\pi L^2}=\left(-\frac{\d}{2}\right)^{1/4}T_c
\ee
We here remark that the above expression for the temperature \eqref{temp} can be
obtained by comparing our solution \eqref{bground} with eqs.(3.6) and (3.7) of \cite{Kim:2007qk}.
However, for this purpose we have to make a coordinate transformation from our $u$ coordinate
to their $z$ coordinate as follows,
\be\label{coord}
u = \frac{L^2}{z}\sqrt{1-\frac{u_3^4z^4}{4L^8}} = \frac{L^2}{z} \sqrt{1-fz^4}
\ee
where $f \equiv u_3^4/(4L^8)$. Note also that the AdS radius $L$ in our solution is the same
as $R$ in their solution. It is not difficult to check that with the above coordinate
transformation our solution \eqref{bground} precisely matches with eqs.(3.6) and (3.7) of
Kim et.al. \cite{Kim:2007qk} solution if we further identify $\d \equiv -2a/f = -8aL^8/u_3^4$.
(Here we like to mention that the solution in \cite{Kim:2007qk} has been obtained by solving the
dilaton-gravity in five dimensions, whereas, our solution is obtained as a particular case
of decoupled geometry of non-susy D3-brane solution which is a genuine type IIB string theory
solution, clarifying the ten dimensional origin of the metric decribed in \cite{Kim:2007qk}.)  
Now since $a$ is related to the temperature as $a = (1/4) \pi^4 T^4$ (given in \cite{Kim:2007qk}),
therefore, we get $\d = -2a/f = (-2\pi^4 L^8/u_3^4) T^4$ and hence \eqref{temp} follows. As the
two solutions are the same, the background \eqref{bground} has a naked singularity at
$u=0$ which corresponds to $z=f^{-1/4}$. One can still define a temperature for this solution
and a detailed discussion on this account has been given in \cite{Kim:2007qk} which we will
not repeat here. However, we mention that there are two mass scales in the corresponding
gauge theory denoted by $a$ and $c$ in ref.\cite{Kim:2007qk} related to the temperature and the
gluon condensate. In our solution they are given as,
\bea\label{ac} & &  a = -\frac{\d \, u_3^4}{8 L^8} = \frac{1}{4} \pi^4 T^4, \qquad c = \frac{u_3^4}{4L^8}
\sqrt{1-\frac{\d^2}{4}},\nn
& & \Rightarrow \quad \sqrt{a^2+c^2} = f = \frac{u_3^4}{4L^8}
\eea
Conversely, the parameters in our solution \eqref{bground} are given in terms of $a$ and $c$ as,
\be\label{u3delta}
\frac{u_3}{\sqrt{2} L^2} = (a^2+c^2)^{\frac{1}{8}} = f^{\frac{1}{4}};\,\,\,
\delta = -\frac{2a}{\sqrt{a^2+c^2}} = -\frac{2a}{f}
\ee
Note from \eqref{ac} that $a=0$ corresponds to $\d=0$ (as $u_3=0$ gives pure AdS, we assume
it to be non-zero) which also implies $T=0$ and so, this is the zero temperature case. In this case $c=u_3^4/(4L^8)$
is related to the zero temperature gluon condensate. On the other hand, $c=0$ corresponds to $\d=-2$,
there is no gluon condensate and from \eqref{temp} it follows that in this case $T = T_c = u_3/(\pi L^2)$.
We notice from \eqref{bground} that for $\d=-2$, the metric reduces to AdS$_5$ black hole (apart from S$^5$ factor)
with no naked singularity and $T_c$ is the corresponding Hawking temperature. This is a complete deconfined phase
(a thermalized state)
and $T_c$ is the critical temperature for the confinement-deconfinement phase transition. Except $\d=-2$, the
metric has a naked singularity for all other values of this parameter and as argued in \cite{Kim:2007qk},
the system in this case is in a non-thermalized state containing free gluons as well as gluon condensate. The
singularity implies the limitation to describe such system by gravity configuration. However, as the action
itself is finite, there must be a concept of temperature (this is given by the relation \eqref{temp}) which
can also be understood from the corresponding gauge theory point of view that gluon condensate can exist
at finite temperature. As the parameter $\d$ varies between the value 0 and $-2$, the quark-gluon plasma goes
from the zero temperature gluon condensate phase to the fully deconfined phase at temperature $T_c$. Here
the variation is studied by hand and not as a time dependent process. Also, note that $\d$ in the gravity
solution \eqref{bground} simply gives an anisotropy in the time direction, whereas, in the gauge theory side this
is related to a combination of temperature and the gluon condensate of the quark-gluon plasma.

To proceed, we note that in the background \eqref{bground}, the axion field is zero, but 
its fluctuation can be non-zero. Now we assume $\chi$ as the fluctuation of 
axion. Thus the corresponding linearized equation of motion in the string frame is 
\be\label{axion1}
\nabla_\text{string}^2\chi=0
\ee
where, $\nabla_\text{string}^2$ is the Laplacian operator in given gravity 
background \eqref{bground} in string frame. Now for simplicity, we assume that the 
axion fluctuation is polarized along the world-volume and symmetric on the 
transverse sphere $S^{5}$. So we can take the ansatz $\chi=\kappa(u)e^{ik_\mu 
x^\mu}$, where $k^\mu=(E,{\vec p})$ is the four-momentum satisfying $E=\sqrt{
m^2+p^2}$ and $x^\mu=(t,x^i)$ is the world-volume coordinates. Now taking this 
ansatz in \eqref{axion1}, we get
\bea
&&\partial_u^2\kappa+\left[\frac{5}{u}+\left(1+\frac{\d_1}{2}\right)\frac{
\partial_uG(u)}{G(u)}\right]\partial_u\kappa\nn
&&\quad +\frac{L^4}{u^4}G(u)^{-\frac{5}{4}+\frac{\d}{8}}\left(E^2G(u)^{
-\frac{\d}{2}}-p^2\right)\kappa=0\nonumber
\eea
where $\d_1=\sqrt{6-\frac{3}{2}\d^2}$. To get a simplified form of the above 
differential equation, we take the coordinate transformation $u=u_3e^y$. Here 
the coordinate range $0<u<\infty$ is transformed to $-\infty<y<\infty$. 
Therefore the near-singularity region is magnified in this new radial coordinate. 
Under this transformation, the above equation becomes, 
\bea
&& \partial_y^2\kappa+\left[4+\left(1+\frac{\d_1}{2}\right)\frac{\partial_yG(y)}
{G(y)}\right]\partial_y\kappa\nn
&&\quad +\frac{L^4}{u_3^2}e^{-2y}G(y)^{-\frac{5}{4}+\frac{\d}{8}}
\left(E^2G(y)^{-\frac{\d}{2}}-p^2\right)\kappa=0\nonumber
\eea
Here, $G(y)=1+e^{-4y}$. Now we replace $f(y)=e^{2y}G(y)^{\half+\frac{\d_1}{4}}
\kappa(y)$ to get the Schr\"odinger-like wave equation,
\be\label{Schroedinger}
\frac{d^2f}{dy^2}-V(y)f=0
\ee
where the potential corresponding to the axion fluctuation is
\bea
&&V(y)=4\frac{1+2e^{-4y}}{(1+e^{-4y})^2}+\frac{6-\frac{3}{2}\d^2}{\left(1+e^{4y}
\right)^2}\nn
&&-\frac{L^4}{u_3^2}e^{-2y}G(y)^{-\frac{5}{4}+\frac{\d}{8}}\left((m^2+p^2)
G(y)^{-\frac{\d}{2}}-p^2\right)\label{potential}
\eea

\begin{figure}
\includegraphics[width=0.45\textwidth,height=0.3\textwidth]{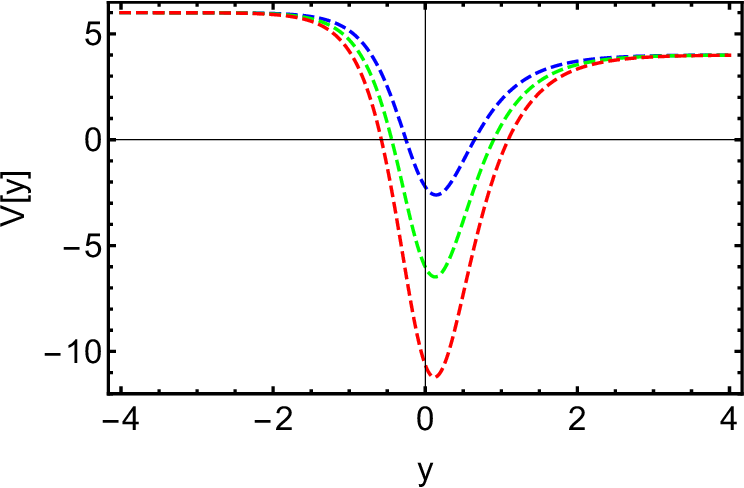}
\caption{The plot of the potential $V(y)$ vs $y$ for $\d=0.0$ and $m=4.0$[blue], 
$5.0$[green] and $6.0$[red]. Here $p$ is taken to be zero.}
\label{fig:pot}
\end{figure}

In Figure \ref{fig:pot}, we have shown the plot of the potential \eqref{potential}
against the dimensionless coordinate $y$. 
When $y$ is very large and negative, the second term of the potential 
dominates and $V(y)$ saturates to $6-\frac{3}{2}\d^2$ which is always positive
and non-zero for the non-constant dilaton. 
On the other hand, at large positive $y$, the first term dominates in 
\eqref{potential} and $V(y)$ merges to a positive constant value which is $4$. 
But in intermediate regime, near $y=0$,  we find a small potential well where $V(y)$ 
is negative. So the expected glueball state is bounded in this negative potential, 
i.e.,the state is confined within this small potential well. Now to evaluate the 
energy of such states with WKB approximation, we need to know the turning points 
in the potential well, i.e., the points between which the associated wave function 
exists. Here we have already seen two turning points where potential 
changes sign. Those points can be found analytically from the asymptotic expansions 
of $V(y)$. The asymptotic forms of this potential at positive and negative 
infinity give the turning points. 

For the positive asymptote, $y\gg 0$,
\be
V(y)\approx 4-\frac{L^4m_a^2}{u_3^2}e^{-2y}
\ee
So the turning point in the positive segment $y_+=\ln(\frac{L^2m}{2u_3})$. 
At the negative asymptote, $y\ll 0$,
\be
V(y)\approx \left(6-\frac{3}{2}\d^2\right)-\frac{L^4}{u_3^2}(m^2+p^2)e^{3(1+\d/2)y}
\ee
The negative turning point is $y_-=\frac{2}{3(2+\d)}\ln\left(\frac{u_3^2}{L^4}
\frac{6-\frac{3}{2}\d^2}{m^2+p^2}\right)$. 
Therefore using WKB approximation in \eqref{Schroedinger}, we get the following equality, 
\be\label{wkb}
\int_{y_-}^{y_+}dy\sqrt{-V(y)}=\left(n+\half\right)\pi
\ee 
Here the integer $n=0,\,1,\,2,\,3,\cdots$ labels the states. It is clear that 
the right hand side of \eqref{wkb} depends on $n$ only. But the integration 
on the left hand side of \eqref{wkb} is a function of $u_3,\,L,\,\d,\,p$ and 
$m$. So once we evaluate the integration, the mass $m$ of a particular 
level $n$ can be expressed as a function of temperature $T$ and momentum $p$. 
But the depth of the potential-well has to be small enough to use the WKB 
approximation. In Figure \ref{fig:pot}, the depth of the potential increases with 
the increasing value of $m$. It shows the depth for $m=4.0$ is the smallest 
and the depth for $m=6.0$ is the largest among the three given values of $p$ and $\d$.
So this method is valid for the 
lower regime of the spectrum. In this approximation method we can evaluate the 
glueball mass for first few energy levels like $m_{-+},\,m^*_{-+},\,m^{**}_{-+}$ etc.,
i.e., for the ground state, the first and the second excited states etc.

Due to the complicated form of the potential function in \eqref{potential} the 
analytic evaluation of the full integration in \eqref{wkb} is not easy. So 
here we will expand the potential function as a power series of $m$. 
According to the parametric condition of the gravity theory, the gauge 
theory temperature is in the range $0\leq T\leq T_c$. Therefore we are in  
the confined state where the glueball dominates in the theory. Due to 
the interaction among themselves, the three-momentum of the glueball is trivial, 
i.e., $E\gg p$ for glueball, which gives $m=\sqrt{E^2-p^2}\approx E$. 
We scale $E$ and $m$ in such a way that $\frac{L^4}{u_3^2}E^2=\omega^2 
\approx \frac{L^4}{u_3^2}m^2=M^2$. So, $y_+=\ln(\frac{\omega}{2})$ and 
$y_-=\frac{2}{3(2+\d)}\ln\left(\frac{6-\frac{3}{2}\d^2}{\omega^2}\right)$. 
Now as $\omega$ is large, i.e., $\omega \gg 1$ we expand the potential function 
\eqref{potential} as a power series in $1/\omega$ and solve the equation 
\eqref{wkb} to find $E$ or $m$. One can easily check that the expansion 
of $\sqrt{-V}$ has the leading order term of $\mathcal O({\omega})$ and the 
other sub-leading terms of $\mathcal O(\omega^{-1})$, $\mathcal O(\omega^{-3})$, $\mathcal O(\omega^{-5})$ and so on. Thus 
considering the terms in \eqref{wkb} upto various powers of $\omega$ the 
analytic expression of $m$ or $\omega$ can be found.

First we consider only the leading $\mathcal O(\omega)$ term in $\sqrt{-V}$ and the sub-leading term up to $\mathcal 
O(\omega^0)$ in \eqref{wkb}. Then from the equality \eqref{wkb} 
the expression of $\omega$ is found to be the following,
\be\label{order1}
\omega^{(1)}=\frac{4\Gamma[\frac{5}{8}+\frac{3}{16}\d]}{\Gamma[\frac{1}{4}]\Gamma
[\frac{3}{8}+\frac{3}{16}\d]}\left[2+\left(n+\half\right)\pi+\frac{4}{
\sqrt{6}}\sqrt{\frac{2-\d}{2+\d}}\right]
\ee
For $\d=0$
\be
\omega^{(1)}=3.47437 + 2.09752 n\nonumber
\ee

Now as we consider all sub-leading terms up to $\mathcal O(\omega^{-1})$ in $\sqrt{-V}$ and with the terms
up to the order of $\omega^{-2}$ in \eqref{wkb}, the WKB approximation formula 
\eqref{wkb} gives the following expression of $\omega$,
\bea
\omega^{(3)} & = & \frac{2\Gamma[\frac{5}{8}+\frac{3}{16}\d]}{\Gamma[\frac{1}{4}]
\Gamma[\frac{3}{8}+\frac{3}{16}\d]}\left[\left\{3+\sqrt{6}\sqrt{\frac{2-\d}
{2+\d}}+\left(n+\half\right)\pi\right\}\right.\nn
&&+\left\{\left(3+\sqrt{6}\sqrt{\frac{2-\d}{2+\d}}+\left(n+\half\right)\pi
\right)^2\right.\nn
&&\left.\left.-\frac{8}{9}\frac{14-9\d}{2-\d}\Gamma[\frac{7}{4}]\Gamma[
\frac{1}{4}]\frac{\Gamma[\frac{5}{8}-\frac{3}{16}\d]}{\Gamma[\frac{5}{8}+
\frac{3}{16}\d]}\frac{\Gamma[\frac{3}{8}+\frac{3}{16}\d]}{\Gamma[\frac{3}{8}-
\frac{3}{16}\d]}\right\}^\frac{1}{2}\right]\label{order3}
\eea
For $\d=0$
\be
\omega^{(3)} = 2.3435 + 1.0487 n+\sqrt{3.1818 + 4.9157 n + 1.0999 n^2}\nonumber
\ee
Here we have found the approximate expressions of glueball energies $\omega^{(1)}$ 
and $\omega^{(3)}$ by considering the sub-leading order terms upto ${\mathcal O}(\omega^0)$ 
and ${\mathcal O}(\omega^{-2})$ respectively, starting from the leading order ${\mathcal O}(\omega)$ 
on the left-hand-side of \eqref{wkb}. Now the energy or mass ratio of the first 
exited state to the ground state are $\omega^{(1)*}/\omega^{(1)}=1.60371$ and 
$\omega^{(3)*}/\omega^{(3)}=1.5567$. The Lattice calculation has found this ratio 
$1.4602$ \cite{Braga:2017apr}, where $m_{-+}=2.477$GeV and $m_{-+}^*=3.617$GeV. 
So for the second approximation $\omega^{(3)}$, the holographic result gets better in 
accuracy. Now we consider the next sub-leading order of $\omega$ in $\sqrt{-V}$ and all of the sub-leading terms 
with orders higher than ${\mathcal O}(\omega^{-4})$ in \eqref{wkb}. In this order, the the energy is denoted $\omega^{(5)}$. For, zero 
temperature $\d=0$, $\omega^{(5)}$ can be found from the following equation.
\bea
&&-5.60086-\frac{0.804218}{\omega^{(5)8/3}}+\frac{14.8333}{\omega^{(5)4}}+\frac{1.28182}{\omega^{(5)3}}\nn
&& \quad +1.49776 \omega^{(5)}+\frac{3.46074}{\omega^{(5)}}=\left(n+\frac{1}{2}\right)\pi
\eea
In the Table \ref{table1}, we have given 
the calculated values of $\omega^{(1)},\,\omega^{(3)},\,\omega^{(5)}$ respectively at $\d=0$. 
There, in the table, we have also calculated the spectrum numerically. In the 
numerical estimation, we have considered the whole $\sqrt{-V}$ in \eqref{wkb}, 
without any approximation. The turning points have been found by the numerical root 
finding method and using those roots the numerical integration has been done to 
solve the equation for $\omega$. Thus, unlike the analytical solutions, the 
numerical solution of $\omega$ includes no approximation on the value of $\omega$. 
Therefore we argue that the 
numerical solution is the exact solution of \eqref{wkb} in our case, as the full 
analytical solution has not been found. We can see, as the more higher order terms 
(order of $1/\omega$) are considered in the analytic solutions, the mass value gradually 
approaches to the exact (numerical) values. In the numerical result $M^*/M=1.5176$, 
whereas analytically $\omega^{(5)*}/\omega^{(5)}=1.5511$. So this ratio also 
approaches to the exact numerical result.

\begin{table}
\begin{center}
\caption{Here the pseudo-scalar glueball energies $\omega$ are given in units of 
$u_3/L^2$ for $\d=0$ for various orders analytically and for the whole series numerically.}
\begin{tabular}{||c|c|c|c|c||}
\hline\hline
$n$ & $\omega^{(1)}$ & $\omega^{(3)}$ & $\omega^{(5)}$ & Numerical\\
\hline
$0$	&	$3.4743$	&	$4.1273$	&	$4.20766$	&	$4.3606$ \\
$1$	&	$5.5719$	&	$6.4251$	&	$6.52683$	&	$6.6177$ \\
$2$	&	$7.6694$	&	$8.6139$	&	$8.71688$	&	$8.7940$ \\
$3$	&	$9.7669$	&	$10.765$	&	$10.8678$	&	$10.942$ \\
$4$	&	$11.864$	&	$12.898$	&	$13.0004$	&	$13.068$ \\
$5$	&	$13.962$	&	$15.021$	&	$15.1230$	&	$15.185$ \\
\hline\hline
\end{tabular}\label{table1}
\end{center}
\end{table}

\begin{figure}
\includegraphics[width=0.5\textwidth,height=0.3\textwidth]{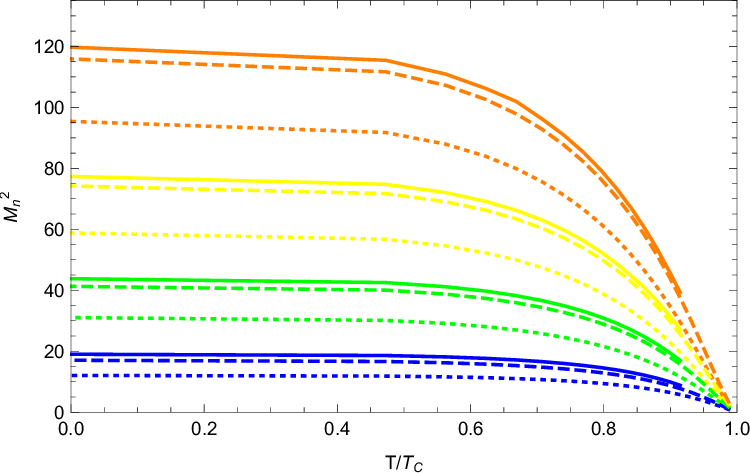}
\caption{ The variation of pseudo-scalar glueball $\text{mass}^2$ or 
$\text{energy}^2$ (we have consider momentum is trivial) in the unit 
of $u_3^2/L^4$ with temperature $\frac{T}{T_C}=\left(-\frac{\d}{2}
\right)^\frac{1}{4}$ is shown for different states [$n=0$ (blue), 
$n=1$ (green), $n=2$ (yellow) and $n=3$ (orange)] for different 
calculations [dotted line, dashed line and solid line indicate the 
analytic results $\omega^{(1)},\,\omega^{(3)}$ and the numerical result 
respectively]. }
\label{fig1}
\end{figure}

Next, we go to the temperature variation of the spectrum. In our analytic results, 
we have seen that the glueball energy explicitly depends on the gravity parameters 
$\d$ and $u_3$ in \eqref{order1} and \eqref{order3}. So, using \eqref{temp}, 
one can write down the glueball energy as a function of $T$. Now if we vary 
$T/T_c$, where $T_c$ is the transition temperature from confined to deconfined 
phase, we can see that $\omega^2 \approx M^2$ decreases with increasing temperature and 
vanishes at $T=T_c$. It is shown in Figure\ref{fig1}. Here we have shown 
the variations for the first four states of the spectrum. For the analytic 
solution \eqref{order1}, the square of the corresponding glueball energy 
$(\omega^{(1)})^2$ is plotted in dotted line and the square of the energy 
$(\omega^{(3)})^2$ given in \eqref{order3} is represented in the dashed line. 
The numerical solution of \eqref{wkb} has been plotted in solid line. Due to 
the limitations of the numerical calculation (probably due to the saddle point 
oscillation in the root finding method) for lower masses, the plot is terminated 
little before $T_c$. Here the plot has been terminated at $0.91T_c$. But in analytic 
evaluation we don't have such limitations and the masses smoothly approach to zero.
It is seen that the glueball 
energy or mass is maximum at $T=0$. As the temperature $T$ increases, initially 
the mass decreases very slowly with $T$. But when the temperature is close to $T_c$, the 
mass decreases very rapidly and becomes zero at $T_c$. The analytical mass 
$\omega^{(1)}$ deviates by large amount from the numerical mass, whereas the 
higher order analytical mass $\omega^{(3)}$ deviates by a small value. 
Also these deviations increase for higher states. It is because, as 
we have previously discussed, the WKB approximation becomes worse for higher 
excited states.

In the above calculations, we have, both analytically and numerically found the 
scaled value of the pseudo-scalar glueball mass, i.e., the mass has been determined 
in units of $u_3/L^2$, where $u_3$ is the fixed mass scale of the gauge theory and 
$L^2=\sqrt{\lambda}=\sqrt{g_\text{YM}^2N}$ is related to the gauge coupling. But 
in reality, according to various high energy studies, the pseudo-scalar glueball 
mass is expected to be on the order of a few GeV. In one of the recent article on 
holographic QCD \cite{Braga:2017apr}, these values at $T/T_c=0$ are $2.477$GeV 
for the ground state and $3.617$GeV for the first excited state. In some of the Lattice 
calculations, the ground state mass is $2.590$GeV\cite{Morningstar:1999rf} or 
$2.560$GeV\cite{Chen:2005mg}, whereas the first excited state mass is $3.640$GeV
\cite{Morningstar:1999rf}. Now if we match our numerical result with these Lattice 
values we find the value of $u_3/L^2$ to be $593.95\text{MeV}$ according to 
\cite{Morningstar:1999rf} and $587.07\text{MeV}$ according to \cite{Chen:2005mg}. 
Again according to the definition \eqref{temp}, the transition temperature $T_c$ are 
$189$MeV and $186.96$MeV for two above values of $u_3/L^2$ respectively. However in 
real QCD, the confinement to deconfinement transition does not occur at a particular 
temperature. The transition occurs through a range of temperature $150-200$MeV. Now, 
using $u_3/L^2=587.07$MeV, we have written the masses of $0^{-+}$ at some finite $T$ 
values in Table \ref{table2}. So both in the graph and in this table we can see a 
clear decrement of the glueball masses with increasing temperature. Again, if we 
take the cross-over point at $T_c=175$MeV, $u_3/L^2=549.5$MeV, the glueball 
spectrum is given in Table \ref{table3}. The zero temperature 
glueball masses from this table are $m_{-+}=2.396$GeV and $m_{-+}^*=
3.636$GeV which almost exactly match with the recent holographic result at zero 
temperature \cite{Braga:2017apr}.
\begin{table}
\begin{center}
\caption{Here the pseudo-scalar glueball mass (in MeV) are given for first two states at 
different temperatures, using holographic QCD. Here we have used $u_3/L^2=587.07$MeV.}

\begin{tabular}{|c|c|c|c|c|c|}
\hline\hline
 & $T=0$ & $T=0.25T_c$ & $T=0.50T_c$ & $T=0.75T_c$ & $T=0.90T_c$\\
\hline
$m_{-+}$ & $2559.95$ & $2557.8$ & $2523.93$ & $2332.18$ & $1846.11$\\
$m_{-+}^*$ & $3885.07$ & $3880.73$ & $3812.81$ & $3440.24$ & $2584.08$\\ 
\hline\hline
\end{tabular}
\label{table2}
\end{center}
\end{table}

\begin{table}
\begin{center}
\caption{Here the pseudo-scalar glueball mass (in MeV) are given for first two states at 
different temperatures, using holographic QCD. Here we have used $T_c=175$MeV. 
So, $u_3/L^2=549.5$MeV.}

\begin{tabular}{|c|c|c|c|c|c|}
\hline\hline
 & $T=0$ & $T=0.25T_c$ & $T=0.50T_c$ & $T=0.75T_c$ & $T=0.90T_c$\\
\hline
$m_{-+}$ & $2396.12$ & $2394.11$ & $2362.41$ & $2182.93$ & $1727.97$\\
$m_{-+}^*$ & $3636.44$ & $3632.38$ & $3568.81$ & $3220.08$ & $2418.71$\\ 
\hline\hline
\end{tabular}
\label{table3}
\end{center}
\end{table}

Here in this article we have found the pseudo-scalar glueball spectrum at the finite 
temperature from the type IIB supergravity background. The dual gauge theory on the 
non-susy D$3$ brane is a QCD-like theory, in which the effective gauge coupling changes 
with the energy and there is a fixed energy scale related to the gravity parameters. Before 
taking the decoupling limit the non-susy brane may contain tachyonic field but
the proper decoupling limit makes the dual theory tachyon free \cite{Nayek:2015tta}. 
The parametric conditions in gravity theory describe the dual gauge theory in a particular 
range of temperature, $0\le T\le T_c$, i.e., the confined regime. At $T_c$ the QCD 
enters completely into deconfined phase. Near this transition point, the glueball melts into the free gluons. 
The glueball loses its mass with decreasing effective coupling as a 
result of the increasing temperature. Thus the pseudo-scalar glueball mass is maximum 
at zero temperature and zero at $T_c$. Here in our study, we have found the mass in term of 
$u_3/L^2$. So, evaluating the constant $u_3/L^2$ from the zero temperature ground state 
mass given in Lattice data we have found the first excited state mass $3.8$GeV at $T=0$ 
which is $0.2$GeV larger than the past result \cite{Morningstar:1999rf}. This discrepancy 
arises mostly due to the approximations which have been used here in our calculational method. After all, we have seen 
that the mass has a large finite value on the order of GeV at $T=0$, whereas it is zero at 
$T=T_c$. This type of decrement of the glueball mass with increasing temperature is an 
important proof of the QCD phase transition, i.e., the confinement-deconfinement transition. Using 
the same arguments, here we have calculated the transition temperature. We have found it  
to be $186$MeV, which falls in the expected range of $T_c$. Therefore, along with the 
holographic estimation of the finite temperature $0^{-+}$ spectrum, this work also proves 
the existence of the QCD confinement from the holographic approach of the string theory. In this 
method, we are not able to calculate the scalar glueball due to the naked singularity in gravity theory. 
However, using this holographic approach we can also study the finite temperature spectrum and phase
transitions in QCD$3$.

\vspace{1cm}

\noindent{\it Acknowledgements:} We would like to thank the anonymous referee for raising some relevant points
whose clarification has helped us, we hope, to improve the manuscript.

\end{document}